\input{aipcheck}

\documentclass[
    ,final            
  ]
  {aipproc}

\layoutstyle{6x9}

\begin{document}

\title{The dust SED in the dwarf galaxy NGC 1569: Indications for an altered dust composition?}

\author{U. Lisenfeld}{
 address={Dept. de F\'\i sica Te\'orica  y del Cosmos, Universidad de Granada,
Spain}
}

\author{F. P. Israel}{
  address={Sterrewacht Leiden, Postbus 9513, 2300 RA Leiden, The Netherlands}
}

\author{J. M. Stil}{
  address={Physics Department, Queens's University, Kingston ON K7L 4P1, Canada}
}

\author{A. Sievers}{
 address={IRAM, Avenida Divinia Pastora 7, N.C., 18012  Granada, Spain }
}

\author{M. Haas}{
 address={Astronomisches Institut, Ruhr-Universität Bochum (AIRUB), Universitätsstr. 150,
44780 Bochum, Germany}
}

\begin{abstract}
We discuss the interpretation of the dust SED from the mid-infrared to the
millimeter range of NGC 1569. 
The model developed by
D\'esert et al. \cite{des90} including three dust components (Polyaromatic
Hydrocarbons, Very Small Grains  and big grains) can explain the
data using a realistic
interstellar radiation field and adopting an enhanced abundance 
of VSGs. 
A simple three-temperature model is also able to reproduce the data but
requires a very low dust temperature which is considered to be unlikely in
this low-metallicity starburst galaxy.
The high abundance of Very Small Grains might be due to large grain destruction in supernova
shocks. This possibility is supported by ISO data showing that 
the emission at 14.3  $\mu$m, tracing VSGs, is enhanced with respect to the emission
at 6.7 $\mu$m and 850 $\mu$m in regions of high star  formation.
\end{abstract}

\maketitle


\section{Introduction}

The dust abundance and properties in galaxies are expected to depend on their
metallicity and interstellar radiation field (ISRF).
In fact, the IRAS colours in dwarf galaxies are found to be different
from those in spirals \cite{mel94}
implying that the
low-metallicity environment in these galaxies indeed has an effect on
the dust.
NGC 1569 is a nearby  (2.2 Mpc) dwarf galaxy in a post-starburst phase but still
exhibiting a high star formation rate. 
It is an excellent candidate to study the 
dust amount and properties 
 in a low-metallicity (12+log(O/H)=8.19)
galaxy with a high ISRF.

We observed NGC 1569 at 1200 $\mu$m with MAMBO at the IRAM 30m telescope
and at 450 and 850 $\mu$m with SCUBA at the JCMT between 1998 and 2000.
A detailed description of the observations, data reduction and 
modelling is given in Lisenfeld et al. \cite{lis02}.
Here we give a summary of the observations and modelling,
compare them to more recent results and include a discussion
of the spatial variations of the dust SED.

\section{The data}

We determined total galaxy flux-densities from the maps
at 450, 850 and 1200 $\mu$m
by integrating them over
increasingly larger areas until the cumulative flux-densities converged to
a final value. However, it was found that low-level emission from
NGC 1569 extended over most of the SCUBA field of view of 2 arcmin
due to the relatively large size of NGC 1569.
This low-level emission made a reliable determination
of the zero-level of the maps impossible. Fortunately, this was not
a problem with the larger IRAM 1200 $\mu$m field.
Therefore, we solved the problem by fitting the 850 and 450 $\mu$m cumulative growthcurve
to the scaled growthcurve at 1200 $\mu$m, increasing the total fluxes at 850 and 450 $\mu$m
by about 30\% in this way.

The data, together with the IRAS data at 12, 25, 60 and 100 $\mu$m taken from
the IRAS Point Source Catalog, indicate various pecularities in comparison
to spiral galaxies.
The high 25/12 $\mu$m ratio indicates that 
the PAH contribution, dominating the 12 $\mu$m emission of our Galaxy,
 must be low with respect to the  contribution
of Very Small Grains dominating at 25 $\mu$m.
Furthermore, we found a high 25/850 $\mu$m ratio and a relatively flat
spectrum in the (sub)millimeter range.

\section{Modelling the dust SED}

\begin{figure}
\centerline{\includegraphics[height=.3\textheight,angle=0]{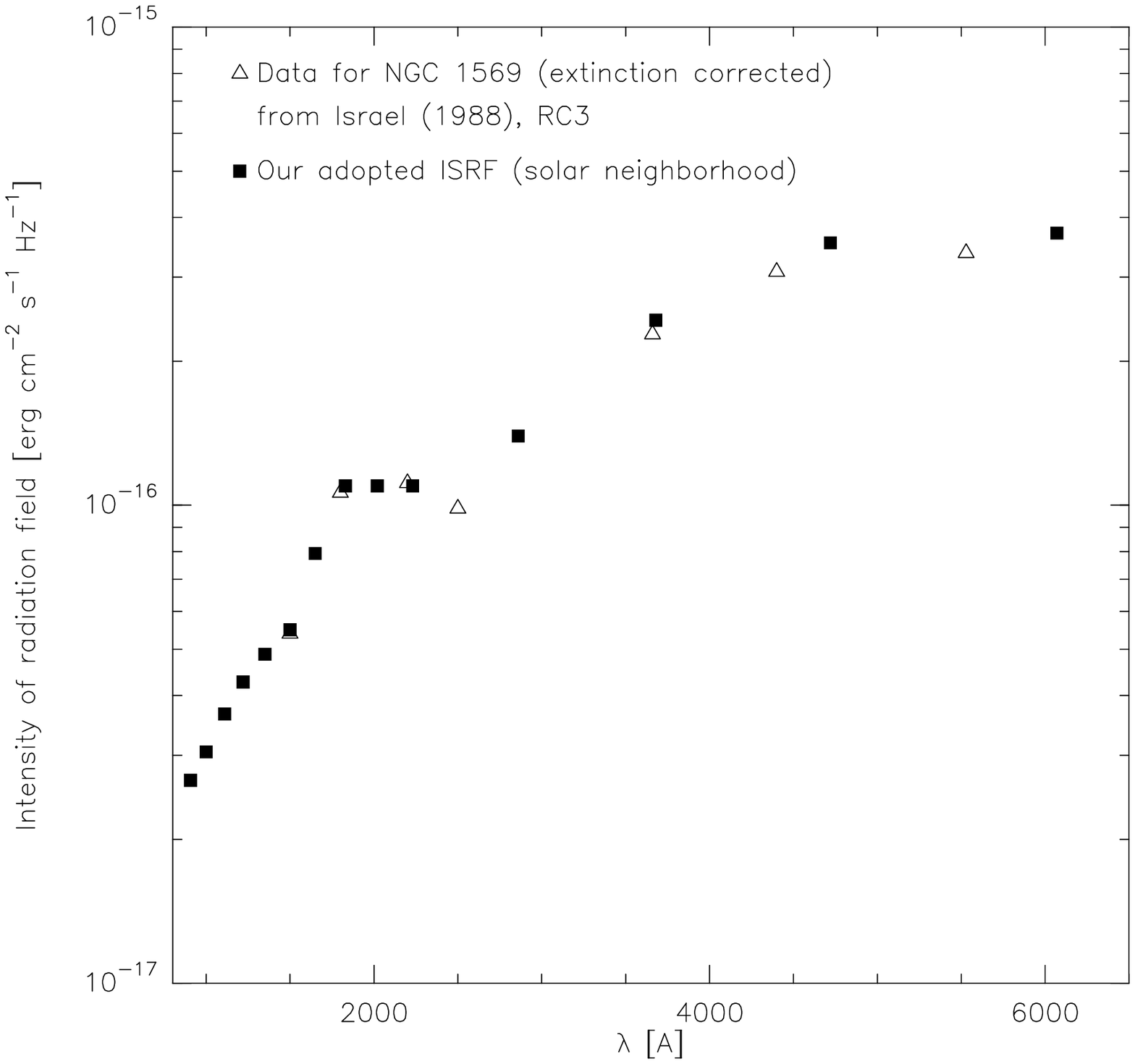}
\includegraphics[height=.3\textheight,angle=0]{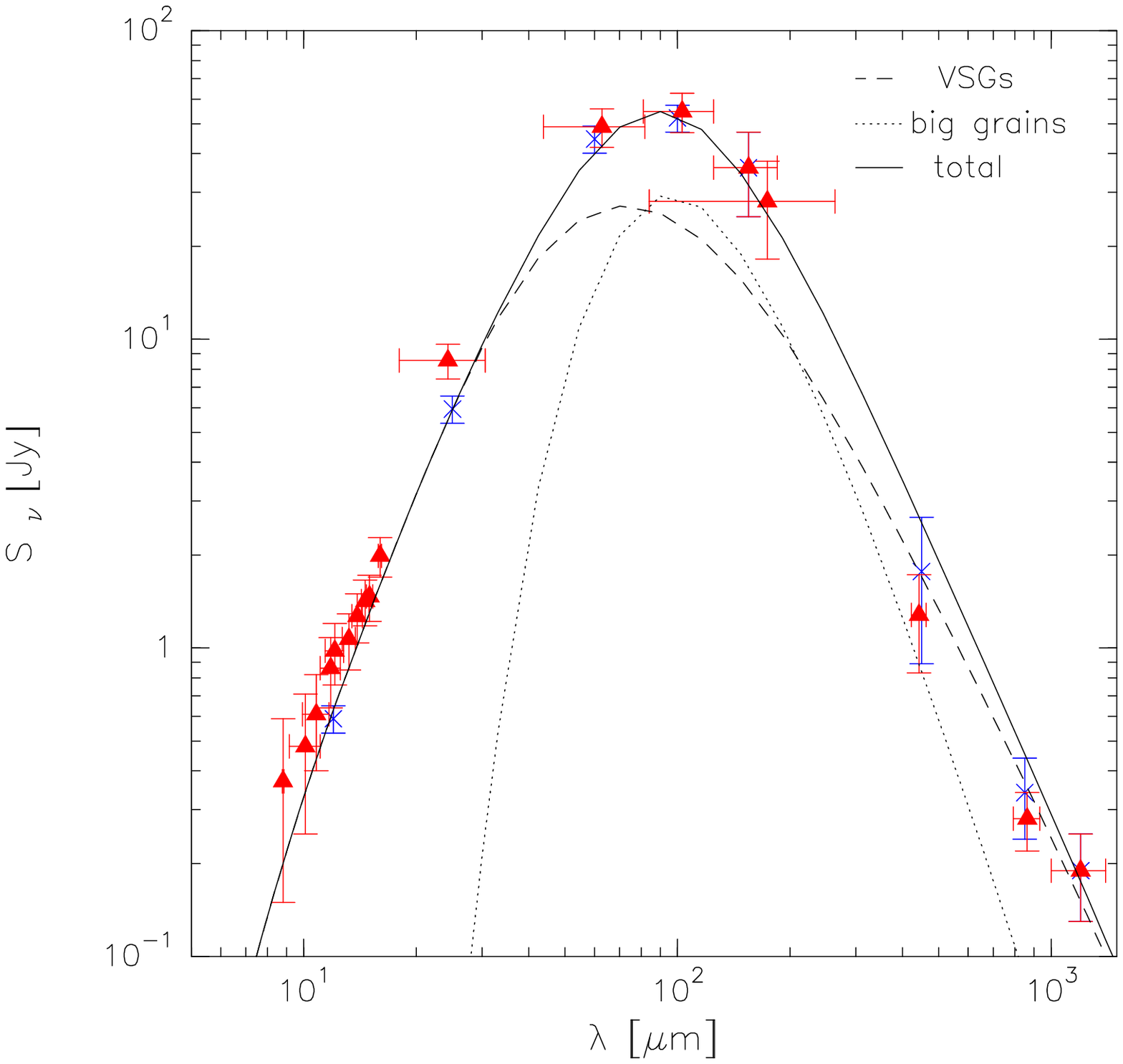}}
  \caption{{\bf Left:} The interstellar radiation field adopted in the
modelling of NGC 1569 (black squares). The open triangles are  
data from Isreal \cite{isr88},
corrected for Galactic foreground extinction 
which represents the bulk of the extinction in NGC 1569.   {\bf Right:}
The best fit based on the model of D\'esert et al. \cite{des90}. The  (blue) 
crosses are the data used in Lisenfeld et al. \cite{lis02} and the (red) triangles the
data of Galliano et al. \cite{gal03}.
}
\end{figure}

We used the model of D\'esert et al. \cite{des90} (hereafter DBP90) in order to fit the data.
This model includes three components: (i) big grains that are in  equilibrium with 
the ISRF, (ii) stochastically heated Very Small Grains (VSGs)
and (iii) Polyaromatic Hydrocarbons (PAHs). 
Since we only have a limited 
number of the data points 
we did not try to exploit the full parameter range of this model
 and only changed the relative contribution of the three components.
We used the ISRF shown in Fig. 1 (left)
derived from ultraviolet to infrared observations of NGC 1569 and
corrected for Galactic extinction which is globally much more important than
intrinsic extinction in NGC1569.

In Fig. 1 (right) we show the best fit result, together with our (sub)millimeter data
and the IRAS data points.  As stated above, the low 12/25 $\mu$m
ratio indicates a relatively low contribution of PAHs with respect to the VSGs.
Within the limited data, we are not able to constrain the contribution of PAHs any further
and therefore do not include them in the fit here. The contribution of VSGs with respect
to big grains is enhanced by a factor of 5.3 in this fit with respect to the Galactic value.
Varying the ISRF in a reasonable way, we found that enhancements of the VSGs between
factors of 2 and 7 give good fits to the data.
In our fit the VSGs are responsible both for the mid-infrared and
for the (sub)millimeter emission. Their broad emission spectrum is due to 
(i) the broad temperature range and (ii) the fact that the frequency dependence
of their dust emissivity is assumed to be  $\nu^{-1}$ (DBP90).

In this figure we also include
data by Galliano et al. \cite{gal03}. The agreement is generally good with
the exception of the data point at 60 $\mu$m and the SCUBA points which are 
lower than ours. The difference between the two sets of SCUBA data can be
explained by the aperture correction that we applied.

As an alternative we tested  a multi-temperature model. 
Because of  the limited amount of data points we were only able to constrain
three temperature components. In spite of the simplicity of the model,
the results are expected to give a reasonable upper limit to the cold dust temperature. 
We needed very cold dust temperatures
in order to 
explain the relatively high flux  and the flatness of the
spectrum in the (sub)mm range.
The best fit to the data was achieved with a cold dust temperature of 7 K.
A dust temperature of 11 K still yielded acceptable agreement with the data  within
the error limits.
Such low dust temperatures can only be achieved in clouds with a very high opacity
and a low filling factor \cite{gal03}. However it needs to be tested whether
the overall emission of such clouds in the high ISRF of NGC 1569 is in agreement
with the observed dust SED of NGC 1569.

\section{Spatial variations of the dust SED}

\begin{figure}
\centerline{\includegraphics[height=.3\textheight,angle=0]{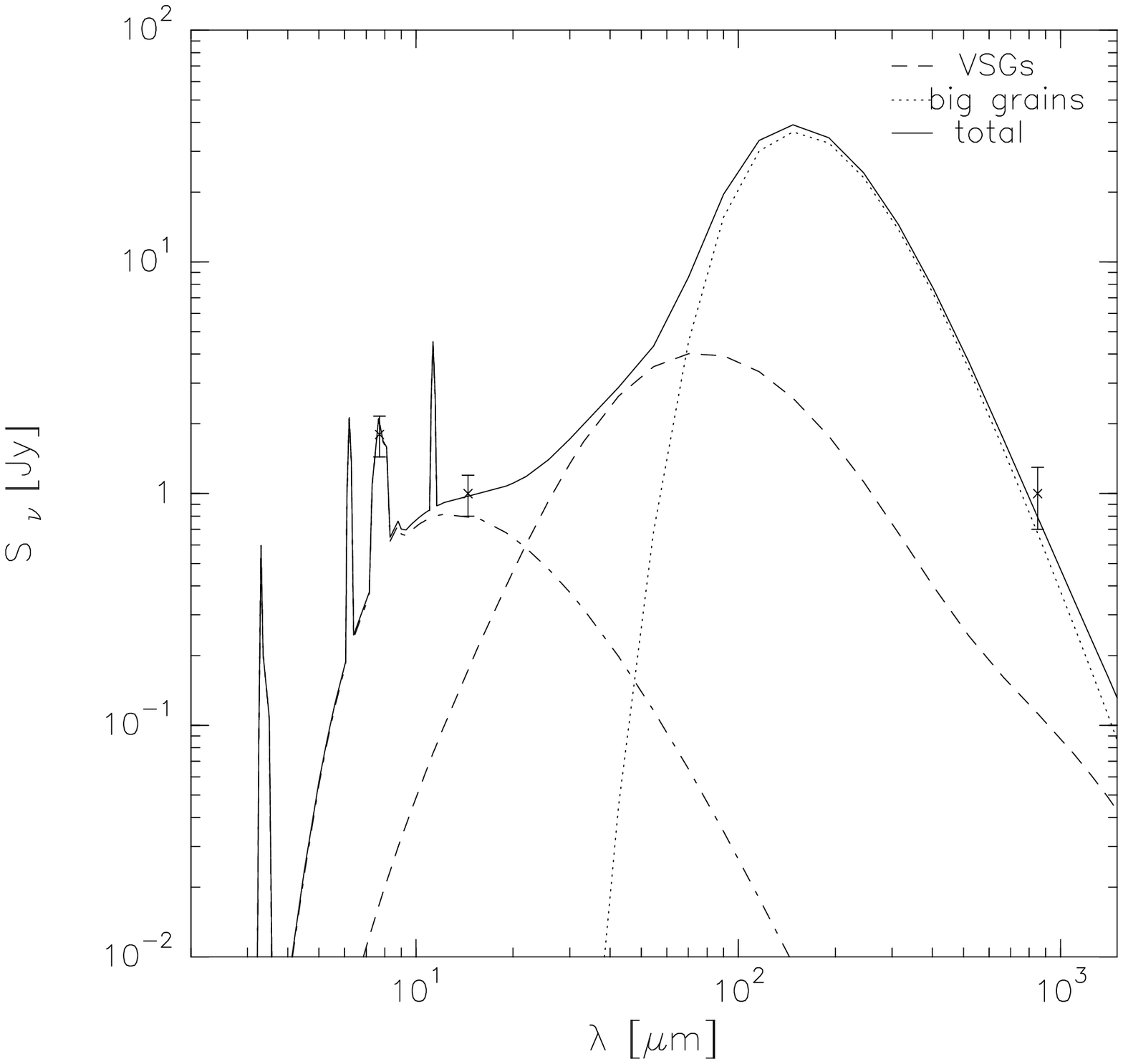}
\includegraphics[height=.3\textheight,angle=0]{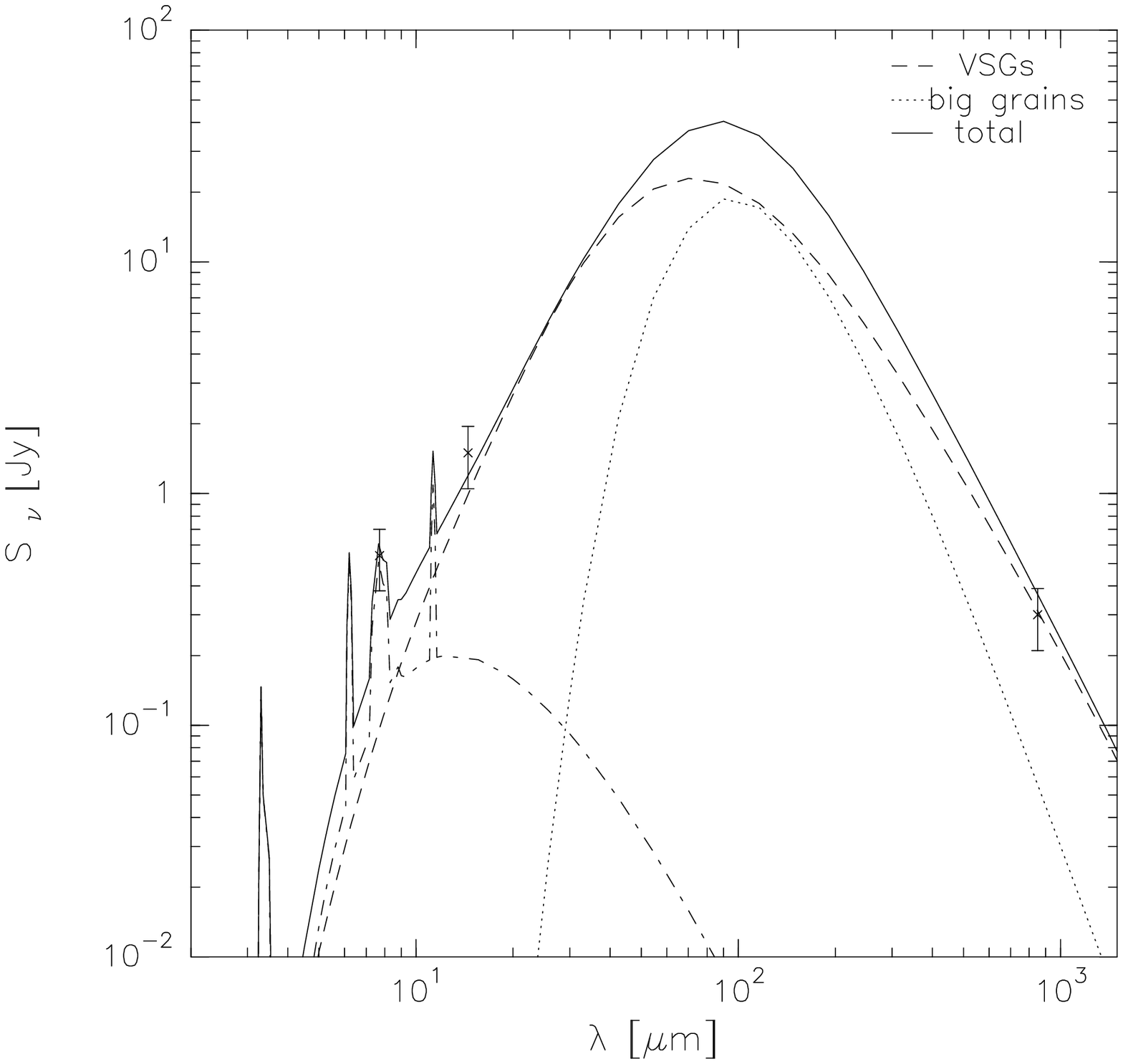}}
  \caption{{\bf Left:} Fit of the model of DBP90 for
an ISRF 20 times lower than the field in NGC 1569 and for  relative abundances
of PAHs, VSGs and big grains as in the solar neighborhood. 
The crosses refer to the data for  low emission regions in NGC 1569
given in Haas et al. \cite{haa00}.  {\bf Right:} Fit of the model of DBP90 for
the ISRF as in  NGC 1569 (Fig. 1, left). The relative abundances in comparison to the
solar neighborhood 
are  3 times lower for PAHs and  6.5 times higher for VSGs. 
The crosses refer to  data for high emission regions in NGC 1569
given in \cite{haa00} , where we have adopted a relative enhancement of the
14.3$\mu$m emission of a factor of 5. 
The absolute flux scales are  arbitrary.
}
\end{figure}

Haas et al. \cite{haa00} compared ISO images at 6.7$\mu$m and 14.3$\mu$m to
SCUBA images at 850 $\mu$m. They found that  the emission at 14.3$\mu$m in 
NGC 1569 is enhanced with
respect to 6.7$\mu$m and 850$\mu$m in
regions of high dust emission, corresponding to regions of high star formation. 
 A similar enhancement was found in the nuclear region of NGC 6946 and in 
the starbursting overlap region in Arp 244. 
In spiral galaxies like NGC 891 or
NGC 7331, on the other hand, there is a very close correlation between the emissions
at 6.7$\mu$m, 14.3$\mu$m and 850$\mu$m.

This indicates that there is a relation between the star formation intensity and
the abundance of VSGs which are dominating the emission at 14.3$\mu$m.
A possible interpretation is that the supernova shocks produced by the star formation
destroy the large dust grains by grain-grain collisions \cite{bor95,jon96}.

The comparison of the maps at different wavelengths in \cite{haa00}
 show
that for NGC 1569 the ratio between 6.7$\mu$m and 850$\mu$m is constant with a value of
$1.08 \pm 0.30$ (and the ratio between 7.7$\mu$m, coinciding with a peak in
the PAH emission, and 850$\mu$m is similarly constant with a ratio of  $1.80 \pm 0.60$)
whereas the ratio between 14.3$\mu$m and 6.7$\mu$m is 1 in low emission regions
and up to 10 in high-emission regions. In Fig. 2 we fit the two sets of data points
with the model of DBP90.
The emission in low-emission region is
very similar to the emission of our Galaxy and can be fitted by a low ISRF and solar 
neighborhood abundances of
VSGs and PAHs. In regions where the emission at 14.3$\mu$m is increased
with respect to 6.7$\mu$m, 7.7$\mu$m and 850 $\mu$m,  an  ISRF as in NGC 1569, an enhanced
abundance of VSGs and a decreased abundance of PAHs  are
able to explain the observations.

\begin{theacknowledgments}
UL is partially supported
by the  Spanish MCyT  Grant  AYA 2002-03338 and
Junta de Andalucia (Spain).
\end{theacknowledgments}

\end{document}